\documentclass[aip, apl, 10pt, reprint, superscriptaddress, amsmath, amssymb]{revtex4-1}
\usepackage{graphicx}
\usepackage{dcolumn}
\usepackage{bm}
\usepackage{upgreek}
\usepackage{mathrsfs}
\usepackage{multirow}
\usepackage{fancyhdr} 
\usepackage{hyperref} 
\usepackage{color}
\usepackage{braket}
\usepackage{lmodern}

\begin{document}

\title{\color{blue}{Microscopic calculation of the optical properties and intrinsic losses in the methylammonium lead iodide perovskite system}}

\author{Lars C. Bannow}\email{lars.bannow@physik.uni-marburg.de}%
\affiliation{%
 Department of Physics and Material Sciences Center, Philipps-Universit\"at, 35032 Marburg, Germany
}%
\author{J{\"o}rg Hader}%
\affiliation{%
  College of Optical Sciences, University of Arizona, Tucson, Arizona 85721, USA
}%
\author{Jerome V. Moloney}%
\affiliation{%
  College of Optical Sciences, University of Arizona, Tucson, Arizona 85721, USA
}%
\author{Stephan W. Koch}%
\affiliation{%
 Department of Physics and Material Sciences Center, Philipps-Universit\"at, 35032 Marburg, Germany
}

\date{\today}

\begin{abstract}
For opto-electronic and photo-voltaic applications of perovskites, it is essential to know the optical properties and intrinsic losses of the used materials. A systematic microscopic analysis is presented for the example of methylammonium lead iodide  where density functional theory is used to calculate the electronic band structure as well as the dipole and Coulomb matrix elements. These results serve as input for a many-body quantum approach used to compute the absorption, photoluminescence, and the optical and Auger losses for a wide range of application conditions. To illustrate the theory, the excitonic properties of the material system are investigated and numerical results are presented for typical photo-voltaic operation conditions and for the elevated carrier densities needed for laser operation. 
\end{abstract}


\maketitle

%
%
Perovskite crystals containing organic cations have gained a lot of attention in recent years, especially for solar cell applications. While high efficiency has been reached, the relatively short life-time of perovskite based solar cells currently still constitutes a significant challenge\cite{Song2016,Tenuta2016,Petrus2017}. The perovskites of interest here can be summarized by the chemical formula AMX$_3$ where A needs to be a large compound such as Cs or even better, to increase the stability, an organic cation. For M, typical atoms are Sn or Pb, and X can be one of the halides I, Cl or Br as well as a mixture of these. Usually, the organic cation is methylammonium (MA) which is CH$_3$NH$_3^{+}$ or formamidinium (FA) which is CH(NH$_2$)$_2^{+}$ but there are also other candidates such as azetidinium or arizidinium\cite{Zheng2018}. Most of the perovskite combinations that can be realized with the listed compounds are semiconductors with band gaps ranging from approximately $1.0\,$eV to $3.8\,$eV \cite{Bokdam2016} where the choice of the halide atom has the strongest impact on the actual band gap value. Depending on the temperature and the composition of the respective perovskite, the crystal structure is cubic, tetragonal or orthorhombic. For changing system temperatures, interesting phase transitions have been observed\cite{Poglitsch1987,Onoda-Yamamuro1990,Stoumpos2013,Whitfield2016}. 

In the current study, we investigate the optical properties of the tetragonal phase of methylammonium lead iodide (MAPbI$_3$). This material system has a band gap of approximately $1.64\,$eV at room temperature \cite{Jiang2016} and is one of the most intensively analyzed perovskites for solar cell applications. 
The band structure of MAPbI$_3$ was previously studied using density functional theory (DFT) \cite{Menendez-Proupin2014,Even2015,Tao2017} or the GW approach\cite{Umari2014,Brivio2014,Bokdam2016,Mosconi2016}. It was shown that DFT calculations which neglect the spin-orbit interaction usually yield band gaps close to the experimentally measured values. However, taking spin-orbit interaction into account severely reduces the band gap \cite{Brivio2014,Umari2014,Menendez-Proupin2014,Yin2015,Even2015,Tao2017} and slightly shifts the conduction band minimum away from the $\Gamma$ point of the Brillouin zone such that the optical transitions become somewhat indirect \cite{Brivio2014}. More advanced DFT calculations utilizing the self-consistent PBE0\cite{Menendez-Proupin2014} or LDA-1/2\cite{Tao2017} approach were shown to perform equally well as GW calculations when spin-orbit interaction is included.

There have been several studies investigating the optical properties of MAPbI$_3$ in order to understand the details behind the good energy conversion in these materials. A significant range of values has been reported for the static and the high-frequency dielectric constants\cite{Brivio2014,Sendner2016,Lin2015,Bokdam2016,Even2015,Umari2014,Hirasawa1994,Juarez-Perez2014}. Consequently, since the computed value for the exciton binding energy $E_\text{b}$ depends on the choice of the dielectric constant, also different values for $E_\text{b}$ have been obtained\cite{Hirasawa1994,Lin2015,Miyata2015,DInnocenzo2014a,Saba2014,Bokdam2016}. Furthermore, there is an ongoing discussion to which degree phonons participate in the screening of the Coulomb interaction potential of the charge carriers. In particular, a large phonon contribution results in a severe reduction of the exciton binding energy\cite{Bokdam2016,Lin2015,Miyata2015}, whereas at room temperature, free carriers would predominate which is one possible explanation for the good energy conversion\cite{Lin2015,Miyata2015}. An alternative explanation is given by \citet{Bokdam2016} who predict the formation of a polaronic state with a smaller band gap and reduced probability of exciton formation. The range of currently discussed exciton binding energies is summarized in Ref. \cite{Herz2016}. The fact that the perovskites under investigation have several active LO phonon modes whose eigen-frequencies are currently not unambiguously identified\cite{Sendner2016, Wright2016, Schlipf2018}, further adds to the problem to identify a reliable value of the exciton binding energy.

Under many conditions, the operation of opto-electronic devices are limited by intrinsic carrier loss processes such as spontaneous emission and/or Auger recombination. Several experimental studies analyzed the strength of the spontaneous emission and Auger processes in MAPbI$_3$, yielding spontaneous emission coefficients in the range of $6\cdot 10^{-11}\,$cm$^3$/s to $2\cdot 10^{-9}\,$cm$^3$/s\cite{Manser2014,Wehrenfennig2014a,Guo2015,Trinh2015,Milot2015} and Auger coefficients in the range of $10^{-28}\,$cm$^6$/s to $10^{-27}\,$cm$^6$/s \cite{Wehrenfennig2014a,Guo2015,Trinh2015}. A comparison between the involved bands of the Auger processes in hybrid perovskites and those in III-V semiconductors can be found in Ref.~\cite{Even2015}. As pointed out in Ref.~\cite{Herz2016}, systematic theoretical investigations on the Auger processes in perovskite materials are deemed highly necessary.

In this work, we use an approach that combines DFT\cite{Kohn1965} calculations for the electronic properties of MAPbI$_3$ with a microscopic many-body approach to compute optical properties such as carrier density dependent absorption and photoluminescence spectra as well as the temperature and density dependent spontaneous emission and Auger loss coefficients. This approach has previously been successfully used to study a wide range of III-V semiconductor material systems \cite{Hader2018a,Hader2018b}. Additionally, we calculate the direction dependent exciton binding energies and discuss these results in the context of the current debate.

%
%
The DFT calculations were performed using the \textit{Vienna Ab-initio Simulation Package} (VASP)~\cite{Kresse1993,Kresse1994,Kresse1996,Kresse1996a} with the Projector Augmented-Wave (PAW) pseudopotential method~\cite{Blochl1994,Kresse1999}. In all DFT calculations the $k$-space was sampled with a $4\times 4 \times 3$ Monkhorst-Pack mesh~\cite{Monkhorst1976}. For the lattice optimization the local density approximation (LDA) as parametrized by Perdew and Zunger~\cite{Perdew1981} was utilized, converging the Hellmann-Feynman forces on the atoms below $10^{-3}\,$eV/{\AA} and the energies to an accuracy of $10^{-6}\,$eV. An energy-cutoff for the plane waves of $800\,$eV was chosen. Optimization of the atom positions, cell shape and volume were performed from a perfect tetragonal crystal structure as a starting point. This results in a pseudo-tetragonal lattice with $a = 8.540\,$\AA$\,$, $b = 8.546\,$\AA$\,$ and $c = 12.611\,$\AA$\,$ which are somewhat smaller than the experimental lattice constants\cite{Stoumpos2013} and the lattice vectors are slightly non-orthogonal. We omitted the inclusion of van-der-Waals corrections based on the recently reported poor performance of Van-der-Waals functionals for the description of the perovskite lattice\cite{Bokdam2017}. For the calculation of the band structures and wave functions, an energy-cutoff of $400\,$eV was used, the energies were converged to an accuracy of $10^{-4}\,$eV and spin-orbit coupling was enabled. In addition, the LDA-1/2 method\cite{Ferreira2008} was used as this method was shown to yield band gaps that are close to the ones obtained with GW for perovskites\cite{Tao2017}. The advantage is that the LDA-1/2 method is computational equally demanding to using the LDA. It is based on Slater's half occupation scheme \cite{Slater1972} and aims at an improved inclusion of the electron self-energies. For iodine we find a cutoff $R_\text{CUT}^\text{I} = 3.71\,$a.u. that maximizes the band gap which is similar to the value \citet{Tao2017} report ($3.76\,$a.u.). For lead, we obtained $R_\text{CUT}^\text{Pb} = 2.00\,$a.u. while \citet{Tao2017} found $R_\text{CUT}^\text{Pb} = 2.18\,$a.u. \citet{Yin2015} found that electronic states, the MA$^{+}$ cations contribute to, are far from the band edges. Therefore, and because it was found that applying the LDA-1/2 technique to the cation has a negligible effect\cite{Ferreira2011}, we do not calculate corrected potentials for the C, N and H atoms. For the non-self-consistent calculations, the $k$-path was sampled throughout the first Brillouin zone with $577$ $k$-points in $\Gamma$A direction ($[111]$ direction), $521$ $k$-points in $\Gamma$M direction ($[110]$ direction), $445$ $k$-points in $\Gamma$R direction ($[011]$ direction), $368$ $k$-points in $\Gamma$X direction  ($[010]$ direction) and $249$ $k$-points in $\Gamma$Z direction ($[001]$ direction). This corresponds to approximately one $k$-point for $10^{-3}\,$\AA$^{-1}$.

From the wave functions, the dipole and Coulomb matrix elements are calculated. Together with the band structure these are needed as input for calculating the optical properties. Absorption and gain are calculated using the semiconductor Bloch equations\cite{Lindberg1988}. For this Eqs.~(2.1)-(2.4) of Ref.~\cite{Hader1999} are evaluated:
\begin{eqnarray}\label{eqn:A}
 \frac{\textnormal{d}}{\textnormal{d}t}p^{\nu\lambda}_{\pmb{k}} &= \cfrac{1}{i\hbar}\Big[\sum\limits_{\lambda',\nu'}\big(\epsilon^{\nu,\nu'}_{\pmb{k}} \delta_{\lambda,\lambda'}+\epsilon^{\lambda,\lambda'}_{\pmb{k}}\delta_{\nu,\nu'}) p^{\nu'\lambda'}_{\pmb{k}}\nonumber\\*
 &+\left(1-f^{\lambda}_{\pmb{k}}-f^{\nu}_{\pmb{k}}\right)\Omega^{\lambda\nu}_{\pmb{k}}\Big] + \cfrac{\textnormal{d}}{\textnormal{d}t}p^{\nu\lambda}_{\pmb{k}}\bigg|_{\textnormal{scatt}}\, ,
\end{eqnarray}
where $f_{\pmb{k}}^{\lambda(\nu)}$ is the electron (hole) density and
\begin{eqnarray}
 \epsilon^{\lambda,\lambda'}_{\pmb{k}} &= \varepsilon^{\lambda}_{\pmb{k}}\delta_{\lambda,\lambda'}-\sum\limits_{\lambda'',\pmb{q}}V^{\lambda\lambda''\lambda'\lambda''}_{\pmb{k}-\pmb{q}}f^{\lambda''}_{\pmb{q}},\\
 \epsilon^{\nu,\nu'}_{\pmb{k}} &= \varepsilon^{\nu}_{\pmb{k}}\delta_{\nu,\nu'}-\sum\limits_{\nu'',\pmb{q}}V^{\nu'\nu''\nu\nu''}_{\pmb{k}-\pmb{q}}f^{\nu''}_{\pmb{q}}
\end{eqnarray}
are the electron (superscript $\lambda$) and hole (superscript $\nu$) energies that are re-normalized due to the Coulomb interaction. In these equations the energy dispersion and Coulomb matrix elements obtained from DFT calculations enter in $\varepsilon_{\pmb{k}}$ and $V_{\pmb{k}-\pmb{q}}$, respectively. Furthermore, $\Omega_{\pmb{k}}$ is given by
\begin{eqnarray}
  \Omega^{\lambda\nu}_{\pmb{k}} &= -d^{\lambda\nu}_{\pmb{k}}E(t)-\sum\limits_{\lambda',\nu',\pmb{q}}V^{\lambda\nu'\nu\lambda'}_{\pmb{k}-\pmb{q}}p^{\nu'\lambda'}_{\pmb{q}},
\end{eqnarray}
where the first term describes the interband dipole coupling to the laser field $E(t)$ and the second term describes again a re-normalization caused by the Coulomb interaction $V_{\pmb{k}-\pmb{q}}$. Here, the dipole matrix elements $d_{\pmb{k}}$ and the Coulomb matrix elements $V_{\pmb{k}-\pmb{q}}$ obtained from DFT calculations enter.

In our current study, we investigate Coulomb effects. While phonon-carrier scattering in principle can be included in the scattering term of Eq.~\ref{eqn:A}, this is challenging due to the strong phonon coupling present in the perovskite materials \cite{Bokdam2016} and the multitude of values reported for the dielectric constants and LO phonon frequencies as discussed above. Therefore, the dephasing caused by scattering processes involving phonons and carriers is approximated by a phenomenological dephasing constant of $\gamma = 165\,$meV (equivalent to a dephasing of $25\,$fs) and the scattering term takes the form $-i\gamma p_{\pmb{k}}$. Additionally, we include carrier-carrier scattering in 2nd Born approximation in the scattering term and Coulomb-screening is taken into account whenever Coulomb matrix elements enter the above equations by evaluation of the Lindhard formula.  

For the calculation of the photoluminescence the semiconductor luminescence equations are evaluated which are derived in Ref.~\cite{Kira2006} [Eqs.~(31)-(33) therein].
From these, the spontaneous emission can be calculated by integrating over the luminescence spectra [Eq.~(4) of\cite{Hader2005}]. Finally, the equations yielding the Auger rate can be found in Ref.~\cite{Hader2005} [Eqs.~(9) and (10) therein]. In addition to a summation over the $k$-points, a fourfold sum has to be evaluated such that all possibilities for the initial and final states are considered. The required input for the energy dispersion of each band and the Coulomb matrix elements are taken from the DFT calculations.

More details on the underlying microscopic many-body approach can be found in Ref.~\cite{Haug2009} and references therein. Unless otherwise stated, we use $\varepsilon_\infty = 6.5$ from Ref.~\cite{Hirasawa1994} for the calculation of the optical properties.

%
%
Our DFT calculations yield a $\Gamma$-point band gap of $E_g = 1.48\,$eV which is significantly smaller than the band gap found by \citet{Tao2017} ($E_g = 1.84\,$eV) who also applied the LDA-1/2 method. The reason is that \citet{Tao2017} re-scale their lattice constants to approximate experimental values. To account for the fact that the DFT calculations are at $0\,$K, we extrapolate the temperature dependent band gaps reported in Ref.~\cite{Jiang2016} for the tetragonal phase to $0\,$K finding $E_g(0\,\text{K}) = 1.55\,$eV which is in good agreement with our result. For the microscopic many-body calculations at different temperatures, we shift the conduction bands such that the resulting temperature dependent bandgaps correspond to the experimental values, e.g.  $1.64\,$eV at $300\,$K \cite{Jiang2016}.

\begin{figure}[t!]
\includegraphics[width=8.5cm]{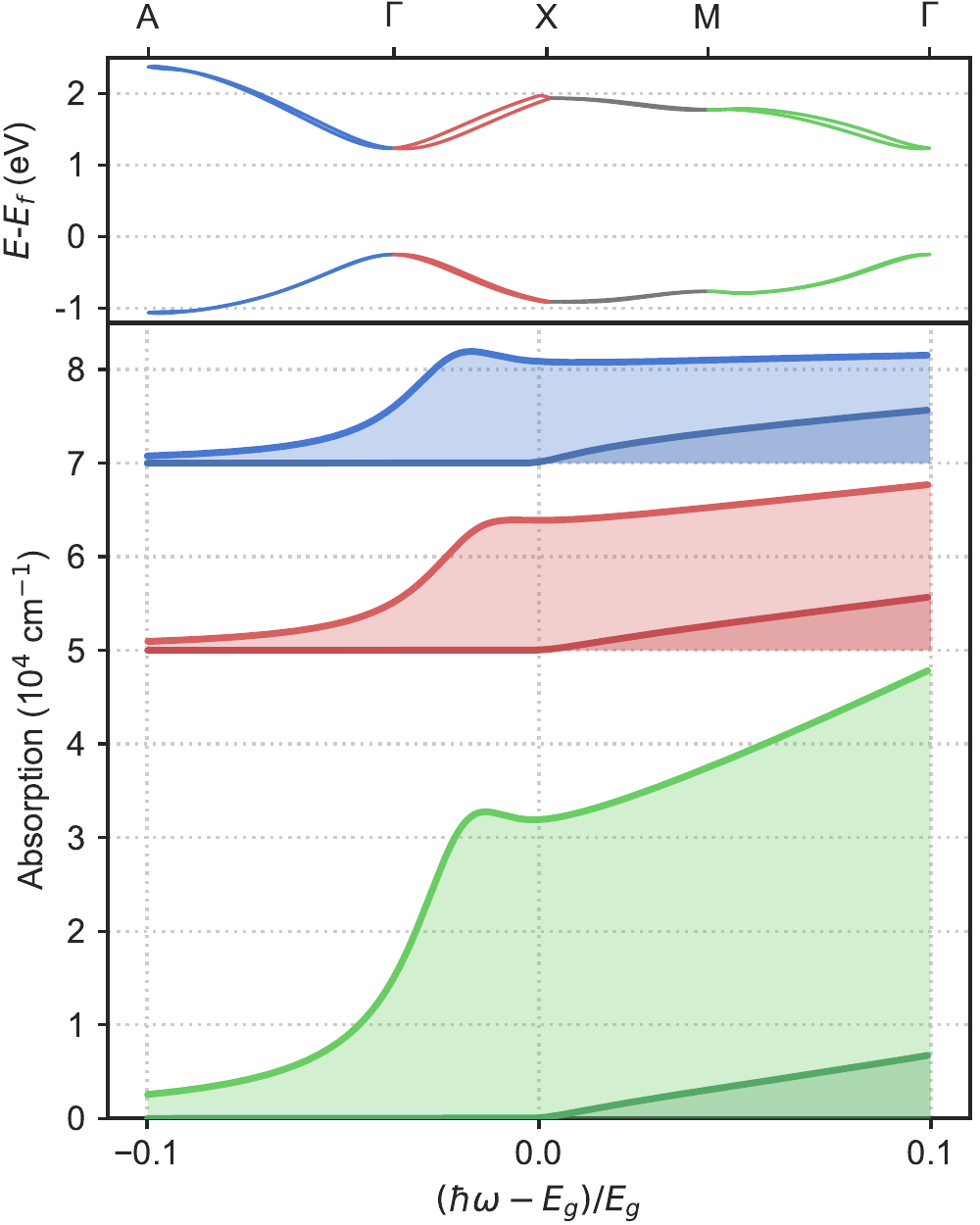}
\caption{The top shows the band structure of MAPbI$_3$ in the tetragonal phase along the A-$\Gamma$-X-M-$\Gamma$ path with the different directions in the same color as the corresponding absorption further down. The bottom shows the absorption for three different directions, namely in the $\Gamma$A (blue), $\Gamma$X (red) and $\Gamma$M (green) direction. The darker lines correspond to the free-particle absorption (no Coulomb interaction). The exciton peak of the $\Gamma$X absorption has been normalized and all other absorption spectra are scaled by the same factor. For better visibility a $y$-offset was applied to the absorption for the $\Gamma$X and $\Gamma$A direction.}
\label{fig:figA}
\end{figure}

The top of Fig.~\ref{fig:figA} shows the obtained band structure of the tetragonal phase of MAPbI$_3$ prior to shifting the conduction bands for the A-$\Gamma$-X-M-$\Gamma$ path and relative to the Fermi level $E_f$. Only the two lowest conduction and two highest valence bands that were taken into account in the absorption calculations are shown. The bottom part of the figure shows the material absorption calculated using the data from  DFT calculations along the $\Gamma$A (blue), $\Gamma$X (red) and $\Gamma$M (green) directions from top to bottom, respectively. The color of the absorption spectra shown in the bottom part of Fig.~\ref{fig:figA} is identical with the color of the band structure for the corresponding direction in the top part of the figure. Both, the absorption spectra including Coulomb interaction and the free-particle absorption spectra have been calculated. For better visibility, a $y$-offset is added to the absorption for the $\Gamma$X and $\Gamma$A direction. An especially strong absorption for the $\Gamma$M direction is evident whereas the absorption for the $\Gamma$A and $\Gamma$X direction are of about equal strength. The actual strength of the absorption depends on the orientation of the crystal in the laser field. According to our results the absorption at the 1s exciton peak is approximately between $1\,\upmu$m$^{-1}$ and $3.5\,\upmu$m$^{-1}$ which agrees with the results summarized by \citet{Shirayama2016} ($2 - 5\,\upmu$m$^{-1}$). Taking more bands into account does not change the absorption in the energy range shown here ($E_g\pm164\,$meV), as further transitions are at higher energies. 

From the onset of the free-particle absorption which corresponds to the band gap $E_g$ and the position of the exciton peak, the exciton binding energy can be computed. The position of the peak and its height depend on the choice of the dephasing constant. For Fig.~\ref{fig:figA}, the dephasing was chosen such that the peak is barely visible. In fact, in experimental measurements around $300\,$K, the exciton peak is absent while it emerges for lower temperatures\cite{Jiang2016,Soufiani2015}. This is most likely due to the reduced phonon scattering at lower temperatures. The position of the maximum can be obtained accurately by using a smaller dephasing constant. It is slightly different for all directions and it depends on the choice of the high-frequency dielectric constant $\varepsilon_\infty$. Several values for $\varepsilon_\infty$ have been reported for MAPbI$_3$ in the literature ranging from $5.0$ to $7.1$\cite{Hirasawa1994,Juarez-Perez2014,Umari2014,Even2015,Sendner2016,Bokdam2016}. For the $\Gamma$M direction we compute an exciton binding energy of $E_\text{b} = 65\,$meV for $\varepsilon_\infty = 5\,$ taken from Ref.~\cite{Sendner2016}, $E_\text{b} = 31\,$meV for $\varepsilon_\infty = 6.5\,$ taken from Ref.~\cite{Hirasawa1994} and $E_\text{b} = 25\,$meV for $\varepsilon_\infty = 7.1\,$ taken from Ref.~\cite{Umari2014}. Table~\ref{Table:A} gives a summary of the obtained exciton binding energies for five different directions in reciprocal space and three different values of $\varepsilon_\infty$. With the exception of the highest value, all are well in the range of the binding energies summarized in Ref.~\cite{Herz2016}. The direction dependence of the exciton binding energy is due to anisotropic effective masses\cite{Umari2014}. 

In the case of CsSnX$_3$ (X = Cl, Br, I), \citet{Huang2013} argue that phonons contribute to the screening and the static dielectric constant $\varepsilon_0$ instead of the high-frequency dielectric constant is to be used. Whether this is the case for MAPbI$_3$ or not is still under debate\cite{Herz2016}. While \citet{Bokdam2016} argue that screening due to phonons can be neglected because the energies of the longitudinal phonon modes in MAPbI$_3$ are by about a factor of $5$ smaller than their calculated $E_\text{b} = 45\,$meV, the authors of Ref.~\cite{Lin2015} claim that screening due to phonons is relevant. The reason for their argument is the large exciton radius of $204\,$\AA$\,$ reported by \citet{Frost2014} which, however, is calculated from a large dielectric constant ($25.7$\cite{Brivio2014}). In Ref.~\cite{Lin2015}, the authors find $E_\text{b} \approx 2\,$meV.

Here, we estimate how our computed exciton binding energies are modified by the inclusion of significant phonon screening. Since for the exciton binding energy $E_\text{b} \propto \varepsilon^{-2}$ applies, our results obtained from using the high-frequency dielectric constant, need to be down-scaled by a factor of $(\varepsilon_0/\varepsilon_\infty)^2$.\cite{Huang2013} The actual magnitude of this factor mostly depends on the choice of $\varepsilon_0$ for which different values are reported in the literature, e.g. $70$~\cite{Lin2015}, $33.5$~\cite{Sendner2016} and $25.7$~\cite{Brivio2014}. Here, we use the values reported in Ref.~\cite{Sendner2016} and find an exciton binding energy of about $1\,$meV. This is close to the value reported by \citet{Lin2015} and even closer to the result of \citet{Frost2014} ($0.7\,$meV).

\begin{table}[b]
    \caption{Exciton binding energy $E_\text{b}$ for different directions in reciprocal space and different $\varepsilon_\infty$ based on Refs.~\cite{Sendner2016,Hirasawa1994,Umari2014}.}\label{Table:A}
    \begin{ruledtabular}
        \begin{tabular}{c c c c c c}
            $\varepsilon_\infty$ & $E_\text{b}$($\Gamma$A) & $E_\text{b}$($\Gamma$M) & $E_\text{b}$($\Gamma$R) & $E_\text{b}$($\Gamma$X) & $E_\text{b}$($\Gamma$Z)  \\
            \hline
            $5.0$&$62\,$meV& $65\,$meV&$51\,$meV&$51\,$meV&  $50\,$meV  \\
            $6.5$&$35\,$meV& $31\,$meV&$28\,$meV&$26\,$meV&  $27\,$meV  \\
            $7.1$&$29\,$meV& $25\,$meV&$23\,$meV&$21\,$meV&  $22\,$meV  \\
        \end{tabular}
    \end{ruledtabular}
\end{table}

\begin{figure}[t!]
\includegraphics[width=8.5cm]{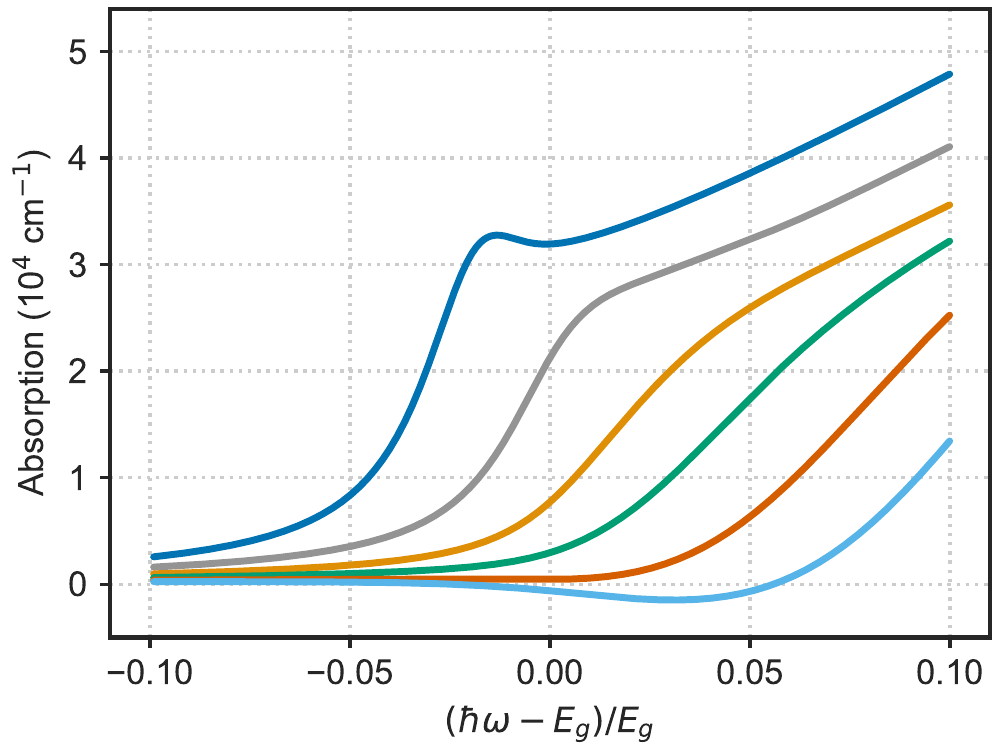}
\caption{Absorption for the carrier densities $n_c = 10^{13}\,$cm$^{-3}$ (dark blue), $n_c = 10^{17}\,$cm$^{-3}$ (grey), $n_c = 7\cdot 10^{17}\,$cm$^{-3}$ (orange), $n_c = 2\cdot 10^{18}\,$cm$^{-3}$ (green), $n_c = 5\cdot 10^{18}\,$cm$^{-3}$ (red) and $n_c = 10\cdot 10^{18}\,$cm$^{-3}$ (bright blue) using the single-particle properties of the $\Gamma$M direction as input for the optical calculations.}
\label{fig:figB}
\end{figure}

Figure \ref{fig:figB} shows the numerical results for different carrier densities $n_c$ using the DFT results in $\Gamma$M direction. We see, that with increasing carrier densities the absorption is shifted to higher energies and around $n_c \approx 10\cdot 10^{18}\,$cm$^{-3}$, we obtain a certain amount of negative absorption, i.e. optical gain.

\begin{figure}[t!]
\includegraphics[width=8.5cm]{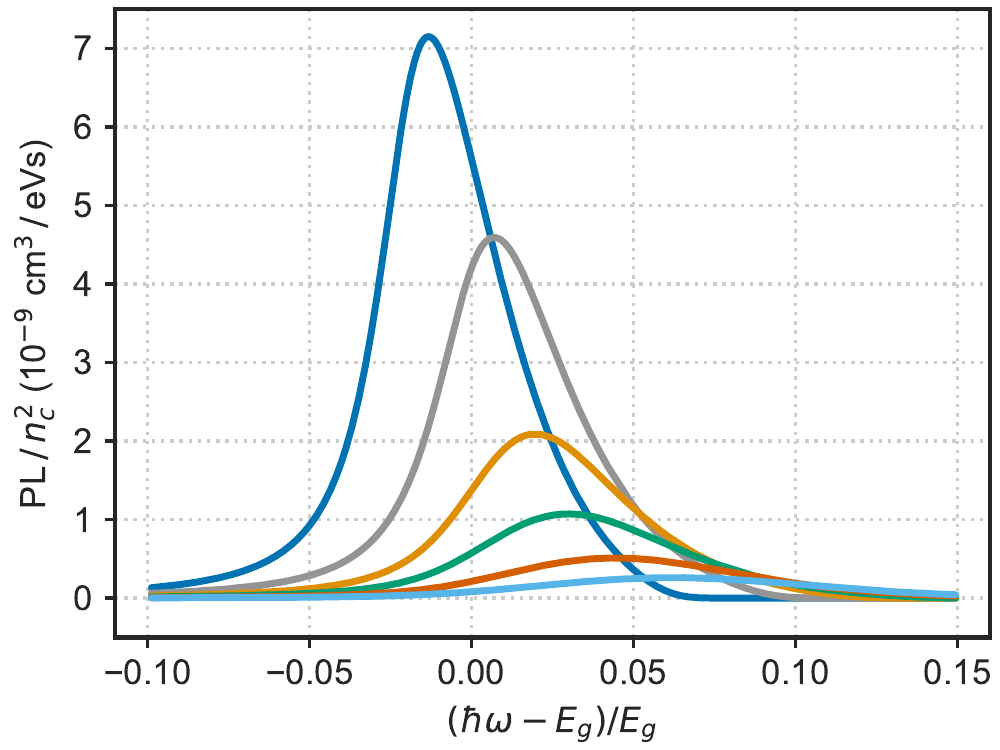}
\caption{Photoluminescence (PL) divided by the square of the carrier densities. Shown are the spectra for the carrier densities $n_c = 10^{13}\,$cm$^{-3}$ (dark blue), $n_c = 10^{17}\,$cm$^{-3}$ (grey), $n_c = 7\cdot 10^{17}\,$cm$^{-3}$ (orange), $n_c = 2\cdot 10^{18}\,$cm$^{-3}$ (green), $n_c = 5\cdot 10^{18}\,$cm$^{-3}$ (red) and $n_c = 10\cdot 10^{18}\,$cm$^{-3}$ (bright blue) using the single-particle properties of the $\Gamma$M direction as input for the optical calculations.}
\label{fig:figC}
\end{figure}

In Fig.~\ref{fig:figC}, we show the computed PL for different carrier densities where the values of the spectra were divided by the square of the respective carrier density for better visibility. For very low carrier densities, we observe that the PL maximum is close to the exciton 1s peak position and, with increasing carrier density, a blueshift is clearly visible. By integrating over the spectra, the rate of carrier loss due to spontaneous emission is obtained and from this, the coefficient $B$ for the spontaneous emission losses is deduced. 

\begin{figure}[t!]
\includegraphics[width=8.5cm]{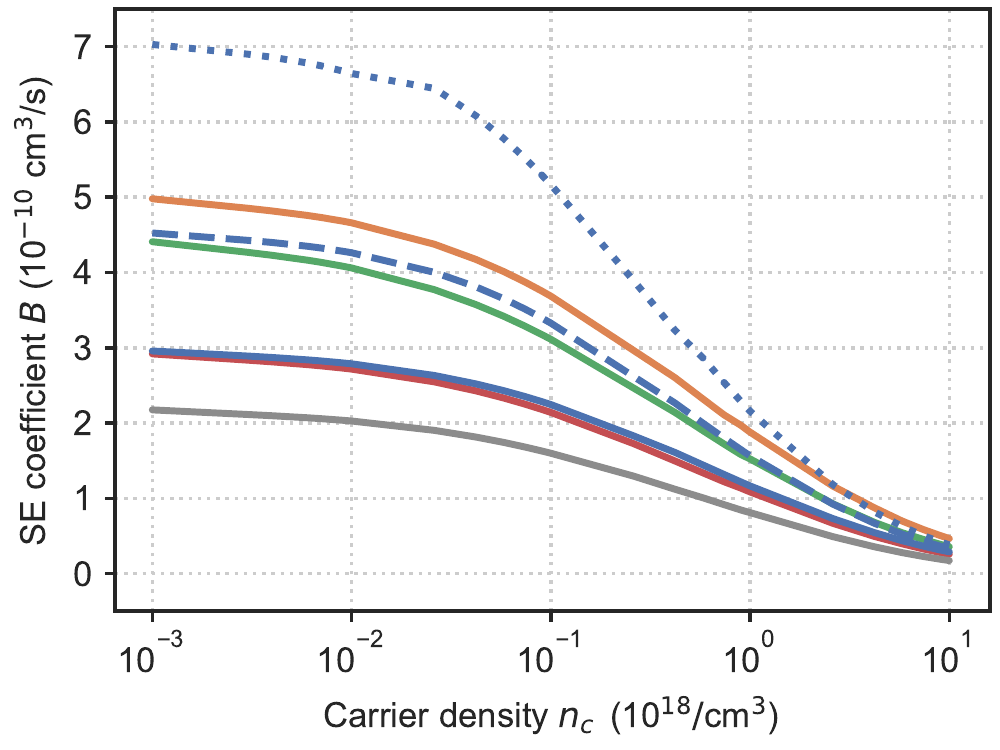}
\caption{The spontaneous emission (SE) coefficient $B$ is shown as a function of the carrier density $n_c$ for three temperatures, namely $T_1 = 170\,$K (dotted line), $T_2 = 235\,$K (dashed line) and $T_3 = 300\,$K (solid line). In the case of $T_3 = 300\,$K, $B$ for the $\Gamma$A (blue), $\Gamma$M (orange), $\Gamma$R (red), $\Gamma$X (green) and $\Gamma$Z (grey) direction is shown.}
\label{fig:figE}
\end{figure}

The obtained $B$ values are plotted in Fig.~\ref{fig:figE} as a function of the carrier density. For the $\Gamma$A direction, $B$ is shown in blue for the temperatures $T_1 = 170\,$K (dotted line), $T_2 = 235\,$K (dashed line) and $T_3 = 300\,$K (solid line). We clearly see that $B$ increases with decreasing temperature, which is in accordance with Ref.\cite{Milot2015}. The reason for this increase can be attributed to the temperature dependence of the occupation probabilities at a given density for states near the bandgap. The computed values for $B$ are also shown in Fig.~\ref{fig:figE}  for the other directions at $300\,$K where it is largest for the $\Gamma$M direction (orange) while it is smallest for the $\Gamma$Z direction (grey). For increasing densities, $B$ decreases due to phase space filling.  Except for very high carrier densities, our values for $B$ are all in the range of those obtained by experimental measurements\cite{Manser2014,Wehrenfennig2014a,Guo2015,Trinh2015,Milot2015} where the reported carrier densities are all well below $10^{19}\,$cm$^{-3}$. 

\begin{figure}[t!]
\includegraphics[width=8.5cm]{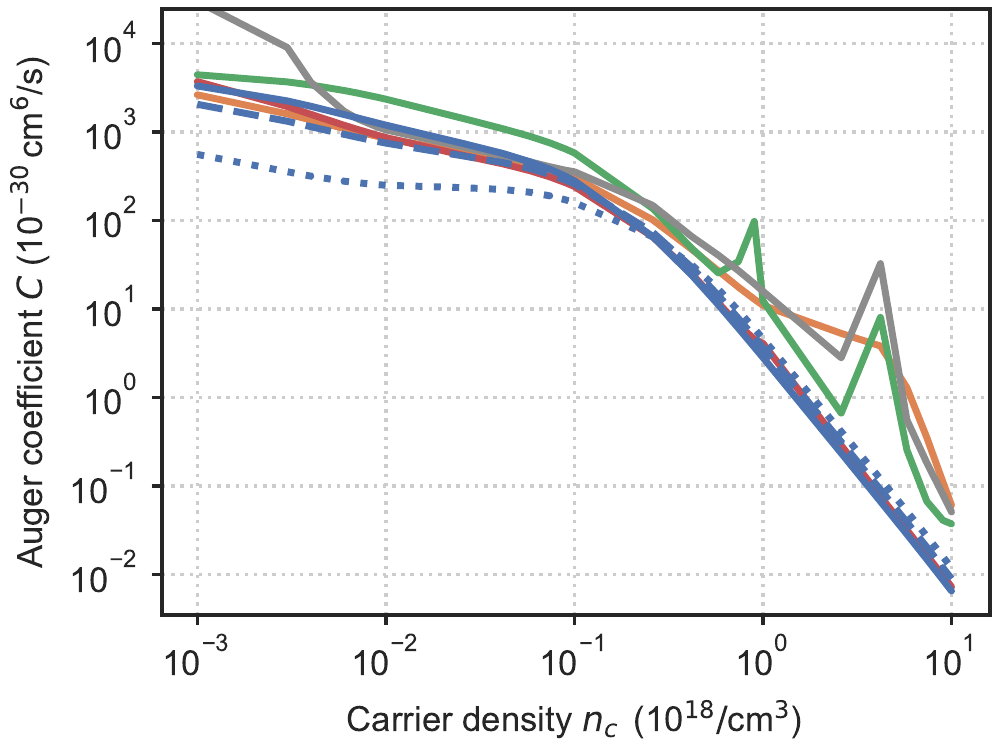}
\caption{The Auger coefficient $C$ is shown as a function of the carrier density $n_c$ for three temperatures, namely $T_1 = 170\,$K (dotted line), $T_2 = 235\,$K (dashed line) and $T_3 = 300\,$K (solid line). In the case of $T_3 = 300\,$K, $C$ for the $\Gamma$A (blue), $\Gamma$M (orange), $\Gamma$R (red), $\Gamma$X (green) and $\Gamma$Z (grey) direction is shown.}
\label{fig:figD}
\end{figure}

In our calculations for $C$, twelve conduction bands and $36$ valence bands were considered which corresponds to $E_\text{CBM}+1.82\,$eV and $E_\text{VBM}-1.83\,$eV at the $\Gamma$-point where $E_\text{CBM}$ is the conduction band minimum and $E_\text{VBM}$ is the valence band maximum. Including more valence bands only slightly increases the Auger coefficient without having any other effects on the results. The next higher conduction bands are energetically too far away to contribute. The complete $k$-path from $\Gamma$ to the edge of the first Brillouin zone is included in the computation of the Auger coefficient. $C$ is obtained for three temperatures, namely $T_1 = 170\,$K, $T_2 = 235\,$K and $T_3 = 300\,$K and plotted as a function of the carrier density $n_c$ in Fig.~\ref{fig:figD}.

With the exception of the $\Gamma$Z direction (grey line), in the low density limit  the Auger coefficients for all directions are very similar. Generally, $C$ in $\Gamma$X direction (green line) is somewhat larger than compared to the other directions. For increasing densities, more and more states are filled and eventually resonances in the Auger coefficient are visible. These resonances occur whenever carriers can make a transition to higher bands without momentum transfer in $k$-space. Since for the five directions considered here, the effective masses differ, these resonances show up at different carrier densities. With increasing temperatures (dotted line $\rightarrow$ dashed line $\rightarrow$ solid line; shown for the $\Gamma$A direction, i.e. in blue), the resonances are shifted to even higher densities (not shown). While at low carrier densities the Auger coefficient increases with temperature, it is almost equal from roughly $n_c \approx 5\cdot 10^{17}/\,$cm$^{3}$ on to higher densities. All in all, these results are in good agreement with experimental findings that reported Auger coefficients in the range of $10^{-28}\,$cm$^6$/s to $10^{-27}\,$cm$^6$/s.\cite{Wehrenfennig2014a,Guo2015,Trinh2015} At carrier densities that are common for solar cell applications on the earth's surface\cite{Herz2016}, we find an Auger coefficient in the range of $1-5\,\cdot\,10^{-27}\,$cm$^6$/s.

Independent of temperature and carrier density, we find that the loss current due to spontaneous emission is usually larger by about one order of magnitude than that due to Auger recombination. An exception are the conditions for which some of the resonances in the Auger rates occur which can enhance the corresponding loss current at that particular carrier density such that it becomes larger than the loss current due to spontaneous emission. 

%
%
In conclusion, we used an \textit{ab initio} based approach combining density functional theory calculations for the electronic properties and microscopic many-body computations to obtain optical properties of MAPbI$_3$. We find a strong direction dependence for the absorption which is largest for the $\Gamma$M direction. Our calculations yield exciton binding energies ($21\,$meV to $65\,$meV) well in the range of reported values that depend on the choice of the high-frequency dielectric constant. Additionally, we compute the reduction of the exciton binding energy in the case when screening due to phonons needs to be accounted for\cite{Lin2015}. Furthermore, we analyze the optical absorption for varying carrier densities. We find a blue shift of the absorption with increasing density and identify the onset of the gain region. This blue shift is also evident in the results of our photoluminescence calculations. Lastly, we investigated the magnitude of the intrinsic loss processes for different temperatures by calculating the spontaneous and Auger recombination coefficients finding a reasonably good agreement with existing experimental data. We extend the experimental results by also calculating the dependence on the carrier density. This is especially strong in the case of the Auger coefficient $C$.  For the complete range of densities used ($10^{15} - 10^{19}\,$cm$^{-3}$) we note that the loss current due to spontaneous emission is by about one order of magnitude larger than the loss current due to Auger recombination, independent of temperature or direction.

%
%
The Marburg work was funded by the DFG via the GRK 1782 "Functionalization of Semiconductors"; computing time from HRZ Marburg, CSC Frankfurt and on the Lichtenberg high performance computer of the TU Darmstadt is acknowledged. The Arizona work was supported by the Air Force Office of Scientific Research under award numbers FA9550-16-1-0088 and FA9550-17-1-0246.

%
%
\bibliography{perovlit2}

\end{document}